# INTERACTION EFFECT DETECTED BY COMPARED OF THE IRREVERSIBLE AND REMANENT INITIAL MAGNETIZATION CURVES IN Ni-Cu-Zn FERRITES

G. GOEV, V. MASHEVA

*Faculty of Physics, "St. Kliment Ohridski" University of Sofia, James Boucher 5, 1164-Sofia, Bulgaria*
*e-mail: gogo@phys.uni-sofia.bg, vmash@phys.uni-sofia.bg.*

Abstract. A new technique for estimation of magnetic interaction effects of initial magnetization curves has been proposed. It deals with remanence, $M_r^{IRM}(H)$, and initial irreversible magnetization, $M_{irr}^i(H)$, curves. The method is applied for single-phase polycrystalline $Ni_{0.85-x}Cu_{0.15}Zn_xFe_2O_4$, (x = 0, 0.2, 0.4 and 0.6), which were synthesized by a standard ceramic technology. A study of the initial reversible and irreversible magnetization processes in ferrite materials was carried out. The field dependence of the irreversible, $M_{irr}^i(H)$ and reversible, $M_{rev}^i(H)$ magnetizations was determined by magnetic losses of minor hysteresis loops obtained from different points of an initial magnetization curve. The influence of Zn-substitutions in Ni-Cu ferrites over irreversible magnetization processes and interactions in magnetic systems has been analyzed.

## 1. INTRODUCTION

The effects of magnetic interactions can change substantially the magnetic characteristic of the materials. These effects are results of different phenomena and there exist several methods of approach them [1]. One of them investigates interaction effects with established $\Delta M$ technique, comparing remanence magnetization, $M_r^{IRM}(H)$, and dc demagnetization remanence, $M_d^{DCD}(H)$ curves [2]. The remanence curves are determined by purely irreversible magnetic changes, but the measurements of the remanences are in zero field.

An initial magnetization curve of a virgin sample can be experimentally obtained by the edges of minor hysteresis loops, plotted in AC magnetic field with progressively increasing amplitude [3]. A method, how the information for irreversible processes of initial magnetization could be derived by the determination of magnetic losses from each minor hysteresis loops, was proposed in Ref. 4. Similar method for irreversible processes of major hysteresis loop can also be applied to Ref. 5. The methods were proved on a Stoner–Wohlfarth model system consisting of disordered non-interacting single-domain uniaxial particles [6] and compared with the results by the remanence curve method [7, 8].

Ni-Cu-Zn ferrites are pertinent magnetic materials for the multiplier chip inductors at high frequencies, which are important components in many electronic devices [9,10]. The magnetic properties of those ferrites are highly sensitive to the sintering conditions. See Refs. 11-14 for more details.

The aim of the present paper is:

(i) to estimate the field dependence of the reversible and irreversible susceptibility and magnetizations by using the initial magnetization curve of the polycrystalline samples, $Ni_{0.85-x}Cu_{0.15}Zn_xFe_2O_4$ (x = 0, 0.2, 0.4 and 0.6),

(ii) to obtain $\Delta M^i(H)$ – plot, describing interaction effects of initial magnetization.

## 2. EXPERIMENT

Zinc substituted Nickel-Copper ferrites of the composition $Ni_{0.85-x}Cu_{0.15}Zn_xFe_2O_4$ (x = 0, 0.2, 0.4, 0.6) were synthesized following the standard ceramic technology described in Ref. 9. The obtained ferrite powders were pressed in ring-forms with outer and inner diameters of 16.5 mm and 10.2 mm respectively, and thickness of 4.6 mm. The final sintering of the rings was carried out at temperature $1125^{\circ}C$ for 4 h in air. The main magnetic parameters such as saturation magnetization, remanence, coercivity and initial permeability at room temperature, measured from the entire hysteresis loops and the scanning electron microscopy (SEM) of the same four samples were reported in Ref. 9. The minor hysteresis loops of toroidal samples were plotted by using a Ricken-Denshi AC B-H Curve Tracer in field with a frequency of 2 kHz at room temperature.

## 3. THE METOD

Both reversible and irreversible processes occur during magnetization along an initial magnetization curve. They can be characterized by the corresponding magnetizations, $M_{rev}^i$ and $M_{irr}^i$, and differential magnetic susceptibilities, $\chi_{rev}^i$ and $\chi_{irr}^i$ respectively. $M_{irr}^i$, $\chi_{irr}^i$, and the energy, $W_{irr}^i$, associated with the irreversible magnetization processes.

The hysteresis losses, $W_{irr,p}^{hyst}(H)$, can be calculated via the hysteresis loop area technique. Where $W_{irr,p}^{hyst}(H)$ is the hysteresis loss for the given "*p*" minor loop. In the present work, the losses of the minor loops of the initial magnetization curve are obtained by using the Fourier decomposition of the curves, as described earlier [15,16].

The irreversible energy, $W_{irr}^i(H)$, susceptibility, $\chi_{irr}^i(H)$, and magnetization, $M_{irr}^i(H)$, are calculated, as it has been shown in Ref. 4.

Remanence magnetizations are equal to $M_r^{IRM}(H)$ and they are obtained from minor hysteresis loops in zero fields.

In macroscopic description, magnetization energy per unit volume of polycrystalline ferromagnetic material consists of many parts. The most important are the exchange energy, which causes the spontaneous magnetization, the magentocrystalline energy, related to the orientation of the magnetization to crystallographic axes, the energy of mechanical strain etc. These energies, together with the magnetic energy in the external field contribute to the irreversible magnetization, $M_{\text{irr}}^i(H)$. When measured in zero field of the remanent magnetization, $M_r^{IRM}(H)$, the energies change. The resulting difference of the magnetizations, $\Delta M^i(H)$ could describe the effects of interaction in the magnetic system:

$$\Delta M^i(H) = M_r^{IRM}(H) - M_{\text{irr}}^i(H). \qquad (1)$$

It is found that the difference, $\Delta M^i(H)$, is equal to zero[4], for Stoner-Wolfarth model system, which is without interaction.

## 4. RESULTS AND DISCISSION

An initial magnetization curve, $M^i(H)$, and minor hysteresis loops, $M_p(H)$, measured for a sample of $Ni_{0.85}Cu_{0.15}Fe_2O_4$ are shown in Fig. 1. For the other investigated samples the curves are similar.

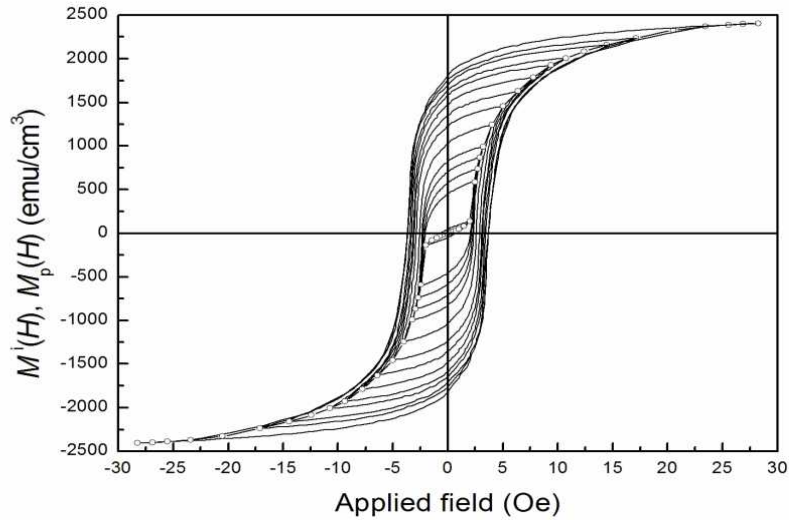

**Figure 1.** An initial magnetization curve, $M^i(H)$ (circles), and minor hysteresis loops, $M_p(H)$, measured for a sample of $Ni_{0.85}Cu_{0.15}Fe_2O_4$.

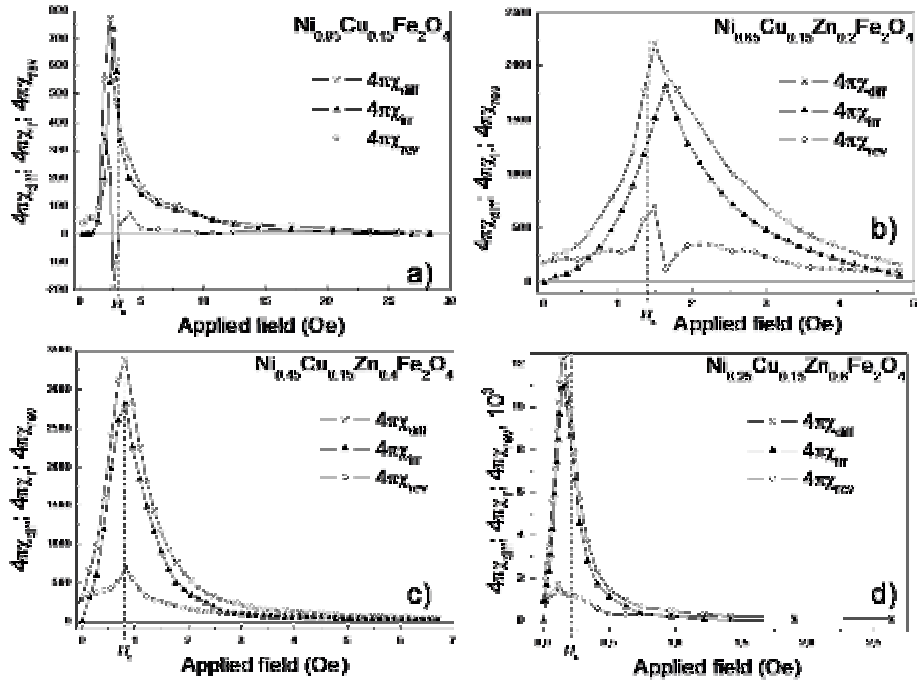

**Figure 2.** Field dependencies of total differential magnetic susceptibility, $\chi^i_{t,\,diff}$ (crosses), irreversible susceptibility, $\chi^i_{irr}$ (triangles) and reversible susceptibility, $\chi^i_{rev}$ (circles) of the initial magnetization curve.

Field dependence of the magnetic susceptibilities, are shown in Fig. 2. The following peculiarities are observed:

(i) The total differential susceptibility and the irreversible susceptibility change almost simultaneously for the samples with Zn-substitution (Fig. 2b, c, d). They have maximum in the region of the coercivity for all samples. The maximum value increases almost linearly with increasing x up to 0.4, but it increases sharply for $x = 0.6$. Almost the same dependence has the initial permeability of the samples [9]. The field of the maximum susceptibility decreases linearly with increasing x.

(ii) The reversible susceptibility has minimum in the region of the coercivity for samples with $x = 0$ and $x = 0.2$ and for $x = 0$ it even changes its sign (Fig. 2a, b). The reversible processes predominate in the fields smaller than coecivity. The irreversible magnetization can sharply increase after then at the expense of the reversible magnetization. The reversible susceptibility can occur with negative values.

(iii) The maximum of total, irreversible and reversible susceptibility and the minimum of only reversible susceptibility are before coercivity ($H_c = 3.2$ Oe) for the sample with $x = 0$ (Fig. 2a). For the sample with $x = 0.2$ the maximum of susceptibilities and minimum of reversible susceptibility are after $H_c$ = 1.4 Oe (Fig. 2b). The coercivity of the sample with Zn concentration x = 0.4 is 0.8 Oe and the maximum of susceptibilities are in same field. The maximum of $\chi_{t,\mathrm{diff}}^i$, $\chi_{\mathrm{irr}}^i$ and $\chi_{\mathrm{rev}}^i$ for x = 0.6 are before coercivity ($H_c$ = 0.2 Oe) (Fig.2d).

The magnetizations obtained for studied Ni-Cu-Zn ferrites are shown in Fig. 3. The reversible magnetization have maximum and minimum for a sample with x = 0 only (Fig. 3a). It increases monotonically after the coercivity. The initial magnetization, the irreversible and reversible magnetization increase monotonically with increasing magnetic field for samples with x = 0.2, 0.4 and 0.6 (Fig.3b, c, d).

On Fig. 3 are shown IRM curves for all samples too. With IRM method separate effects of $H$ and $M_{\mathrm{irr}}^i$ on $M_{\mathrm{rev}}^i$ cannot be directly determined [4].

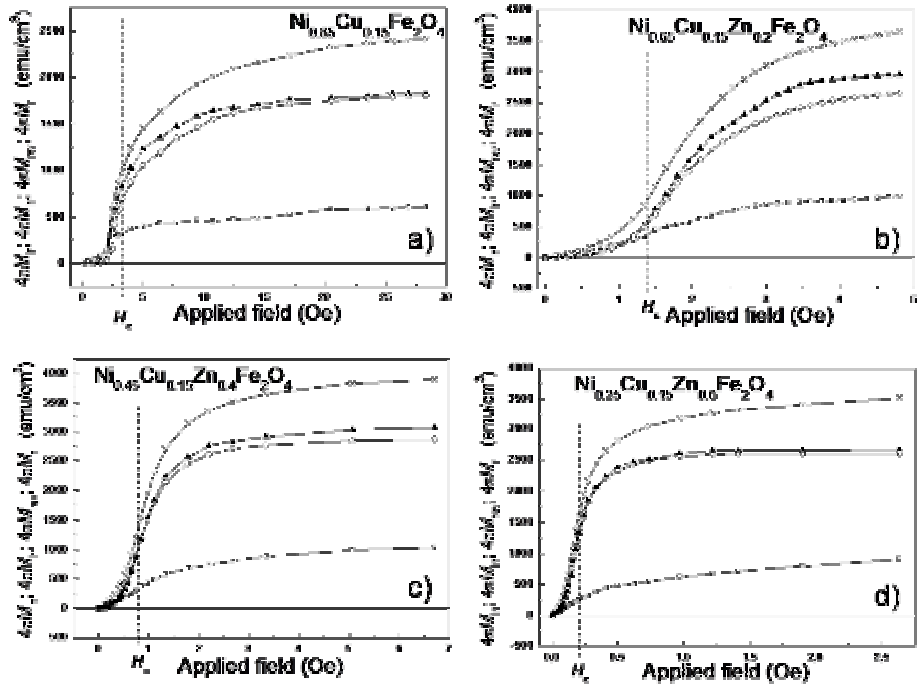

**Figure 3.** Field dependencies of the initial magnetization, $M^i(H)$ (crosses), the irreversible remanent magnetization, $M_r^{IRM}(H)$ (triangles), the irreversible magnetization, $M_{\mathrm{irr}}^i(H)$ (diamonds) and the reversible magnetization, $M_{\mathrm{rev}}^i(H)$ (squires).

The curves of the irreversible magnetization obtained by our method and IRM curves do not coincided. The differences $\Delta M^i(H)$ for the four samples are shown in Fig. 4.

It is seen that interaction effects are exclusively positive for samples with x = 0.0 and 0.2 (Fig.4a,b). Positive interactions assist the magnetization process. $\Delta M^i(H)$- plot has maximum and minimum in the region of the coercivity like reversible magnetization for sample with x = 0 only (Fig.3a, and 4a). For sample with least Zn-substitution x = 0.2, $\Delta M^i(H)$ - plot is insignificant in the region of the coercivity then it increases with applied field.

For sample with x = 0.4 and x = 0.6, $\Delta M^i(H)$ - plot is negative and has minimum before coercivity. In this region interactions hinder magnetization process. After coercivity magnetic interactions increase monotonically and change their sign (Fig. 4c,d). The remanent magnetization, measured in zero field can be smaller or bigger than the irreversible magnetization. The magnetic interactions in a field influence the irreversible magnetization.

On Fig. 4b, 4c and 4d is shown that interactions effect is positive and decrease with increasing Zn-substitution after region of the coercivity.

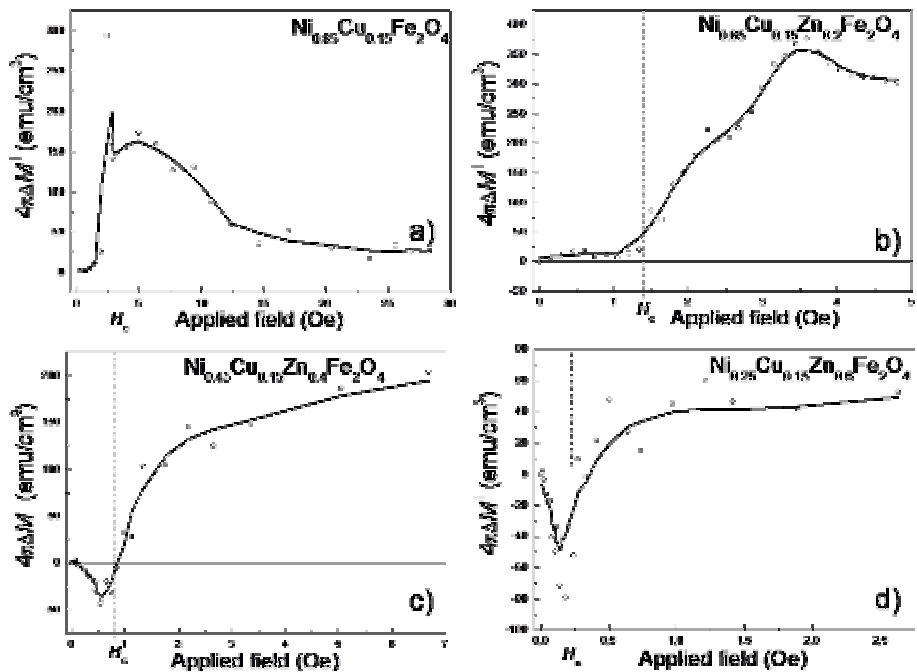

**Figure 4.** Field dependencies of the magnetization, $\Delta M^i(H)$.

## 5. CONCLUSION

The irreversible susceptibilities and magnetizations, the reversible susceptibilities and magnetizations of initial magnetization curves were determined by measuring sets of magnetic losses on minor hysteresis loops for samples of polycrystalline Ni-Cu-Zn ferrites. The method used for the estimation of the irreversible susceptibility of an initial magnetization curve is very sensitive and deals with the energy of magnetization only, and not with the magnetization mechanisms.

The reversible susceptibility has minimum for sample with Zn concentrations, $x = 0.2$ and for sample $Ni_{0.85}Cu_{0.15}Fe_2O_4$ only and even changes its sign.

We have demonstrated that using $\Delta M^i(H)$- plot initial magnetization interactions can be easily quantified. The results obtained show that for sample with Zn concentrations, $x = 0.4$ and $0.6$ interaction effects are negative and positive.